\documentclass[journal,twoside,web]{mcomstd}

\usepackage{cite}
\usepackage{amsmath,amssymb,amsfonts}
\usepackage[mathlines,switch]{lineno}
\usepackage{multicol,multirow}
\usepackage{marginnote}
\usepackage{stfloats}
\usepackage{graphicx}
\usepackage{array}
\usepackage{booktabs}
\usepackage{tabularx}
\usepackage{url}
\usepackage{mcomstd}

\title{Toward a Layer-2 Trigger for AI/ML Lifecycle\\Management in 6G}

\author{{Dharmendra Kumar, \textit{Member,~IEEE}}}

\begin{document}

\maketitle

\begin{firstfootnote}
\raggedright
\normalfont\footnotesize
Dharmendra Kumar is an Independent Researcher in Dallas, TX, USA
(e-mail: dharmendra.sharma4@ieee.org).

This work has been submitted to the IEEE for possible publication.
Copyright may be transferred without notice, after which this version
may no longer be accessible.
\end{firstfootnote}

\bodyfont
\enlargethispage{-8pt}

\begin{abstract}
3GPP has progressively expanded AI/ML lifecycle management in the radio
access network, from one-sided model control to Release~20 support for
two-sided CSI-feedback model pairing. Yet a basic control question remains:
when monitoring detects degradation, how quickly must a corrective action
take effect? To expose this dependency, we stress-test three activation and
rollback strategies in a surrogate regime-shift environment using 150
independently trained PPO candidate policies, each evaluated over 20 matched
noise realizations. We add 0--40 control-step delay only to corrective
lifecycle commands. With no added delay, stability-gated blending reduces
mean post-shift cumulative SLA deficit from 47.02 to 7.57 violation-steps
relative to hard cutover; at 40 steps, the deficit rises to 44.38, only
5.6\% below the hard-cutover baseline. KPI-threshold rollback loses its
advantage within only a few control intervals, while blending degrades more
gradually. These results do not set a physical 6G latency bound. They show
why timing requirements matter for corrective actions: lifecycle performance
depends on when the command takes effect. Motivated by Layer-1/Layer-2 Triggered Mobility,
we examine a standards split in which Layer~3 retains lifecycle
configuration while a compact Layer~2 trigger is considered only for the
latency-critical subset, together with pair-consistency, local-fallback,
and security/freshness requirements.
\end{abstract}

\section{Introduction}
Over three releases, 3GPP has moved AI/ML on the NR air
interface from study to increasingly explicit lifecycle support. Release~18
established the basic vocabulary and collaboration levels~\cite{ref:tr38843}.
Release~19 added one-sided lifecycle functions such as selection, activation,
deactivation, switching, fallback, and performance monitoring~\cite{ref:rel19wid}.
Release~20 extends that work to two-sided models for CSI compression,
including model pairing and inter-vendor interoperability requirements~\cite{ref:rel20wid}.

One question is much less explicit: once monitoring indicates that a deployed
model should no longer be trusted, how quickly must the corrective action
actually take effect?

That question changes the character of lifecycle management. Activation,
switching, and fallback may look like configuration when they are planned in
advance. The same actions become control decisions when they are triggered by
a live KPI, drift detector, or other monitoring signal. In this article,
\emph{rollback} means returning from a candidate to a previously validated
model or to the non-AI fallback state. The value of such an action depends
not only on choosing the right state, but also on reaching it before the
underlying degradation becomes a service-level violation.

Release~19 RRC already carries AI/ML-related capability and configuration
for standardized beam-management and CSI-prediction use cases, including
inference, monitoring, applicability reporting, and data collection~\cite{ref:ts38331}.
The Release~19 work item separately calls for signalling and protocol support
for lifecycle selection, activation, deactivation, switching, and fallback~\cite{ref:rel19wid}.
Together, these specifications establish an important configuration
framework, but they do not attach a bounded lower-latency actuation
requirement to a monitoring-driven corrective action.

This article focuses on that missing timing dimension. We first show why a
monitoring-driven lifecycle procedure is naturally viewed as a closed loop.
We then stress that loop by delaying only the corrective command and measure
how quickly the benefit of rollback and stability-gated blending erodes. The
standards discussion then turns to a precedent already present in
5G-Advanced: Layer-1/Layer-2 Triggered Mobility (LTM), which separates rich
candidate configuration from a compact lower-layer selection trigger. That
pattern provides a useful way to ask which lifecycle events, if any, need a
similar fast path and what consistency and security properties would follow.

The proposal is deliberately limited. It does not replace lifecycle
configuration with a second framework, nor does the experiment prove that
Layer~2 is required. A physical timing budget must first be established and
compared with feasible Layer~3 and lower-layer procedure delays. Consistent
with the lean-and-streamlined 6G principle in RP-250766~\cite{ref:lean}, the
question is whether only the latency-critical trigger should move closer to
the execution point while the richer configuration remains in Layer~3.

\section{How Lifecycle Management Became a Control Problem}
\label{sec:config}
The timing question is easy to miss because the lifecycle framework arrived
incrementally. Each release added useful capability, but no single step
forced a clean separation between slow configuration and time-critical
execution.

Release~18 established the vocabulary~\cite{ref:tr38843}. It defined
collaboration levels between network and device and studied CSI feedback,
beam management, and positioning. It also separated data collection,
training, inference, performance monitoring, and model transfer into
distinct lifecycle concerns. At that stage the work was exploratory, so the
speed of a lifecycle transition was not yet a central specification issue.

Release~19 moved the one-sided case into normative work~\cite{ref:rel19wid}.
Its WID calls for signalling and protocol support for functionality/model
selection, activation, deactivation, switching, fallback, identification,
and performance monitoring. TS~38.331 V19.1.0 contains corresponding
AI/ML-related RRC capability and configuration elements for
inference/prediction, monitoring, applicability reporting, and data
collection~\cite{ref:ts38331}. This makes RRC a natural home for relatively
slow-changing AI/ML configuration; the WID does not, however, state that
every lifecycle trigger must use the same path.

Release~20 adds the two-sided CSI-compression case~\cite{ref:rel20wid}. The
work item includes lifecycle functions, a model-pairing procedure with
identifier and applicability reporting, and inter-vendor interoperability
work based on standardized test components. Timing now has a second
consequence: a delayed transition can affect not only how long a degraded
model remains active, but also whether the two endpoints are using a
consistent model pair.

The key transition occurs when monitoring directly triggers action. A
performance observation, decision rule, signalling path, and model
transition then form a closed loop, regardless of whether the standards text
uses that term. The useful response time is set by how fast the disturbance
develops, not simply by how quickly a configuration procedure can normally
complete.

This distinction separates what is standardized from what is proposed here.
3GPP already includes switching and fallback in the lifecycle scope, but the
WIDs we reviewed do not assign bounded actuation times to those actions.
Release~20 adds model pairing, but not the pair-consistency and recovery
semantics developed later in this article. We treat bounded corrective
actuation, rollback, and coherent paired transition as control requirements
to be studied on top of the standardized lifecycle primitives.

\section{The Missing Latency Budget}
\label{sec:gap}

\subsection{What determines a useful response time}
A monitoring-driven lifecycle loop has four basic elements: observation,
decision, command, and actuation. The observation may be a performance KPI,
an input-distribution statistic, or an inference-confidence proxy. A
detector turns that observation into a decision; the decision is signalled
to the node hosting the model; and the model transition completes the loop.

The useful response time depends on disturbance dynamics, controller cadence,
the violation criterion, and actuation delay. In the regime-shift surrogate
used here, degradation appears as an excursion: the controlled quantity
leaves its operating region and either recovers or continues toward
violation. The intervention window therefore runs from the point at which
the excursion becomes detectable to the point at which corrective action is
too late to prevent or materially reduce the violation. Monitoring,
detection, command transit, and transition execution all consume that
window.

Our experiment isolates one of those terms---command delay. Because the
simulator does not assign a physical duration to a controller step, the
results are reported in control steps rather than milliseconds. A physical
latency requirement should be introduced only after the controller period
and relevant protocol timing assumptions are defined by a system model or
standards source.

\subsection{Where signalling delay enters}
Consider an implementation in which monitoring decides that a candidate
should be switched or rolled back and the corrective action is realized
through an RRC reconfiguration. TS~38.331 specifies the
RRCReconfiguration procedure, but the AI/ML lifecycle WID does not attach an
end-to-end actuation bound to switching or fallback~\cite{ref:ts38331,ref:rel19wid}.
Monitoring and detector delay would therefore be followed by message
preparation, transmission, peer processing, and procedure application. We
do not assume a normative RRC latency in the experiment; instead, we sweep
an abstract additional command delay.

The standards gap is therefore not a claim that any one RRC component is
necessarily too slow. It is that the lifecycle specification does not state
how much end-to-end delay a corrective action may tolerate. Without that
requirement, the procedure definition alone cannot tell a controller whether
its response will arrive inside the useful intervention window.

\subsection{A delay-sensitivity stress test}
The experiment is designed to expose this dependency, not to emulate a
calibrated RAN timeline. The surrogate is a single-cell, two-state control
loop with a KPI-deficit state, an interference proxy, a scalar action, and
linear-Gaussian dynamics that switch from regime A to B at model activation.
The trusted incumbent and 150 candidate controllers are PPO policies trained
with a Lagrangian constraint wrapper; the candidate set intentionally spans
a broad quality range.

\begin{figure*}[!t]
\centering
\includegraphics[width=0.78\textwidth]{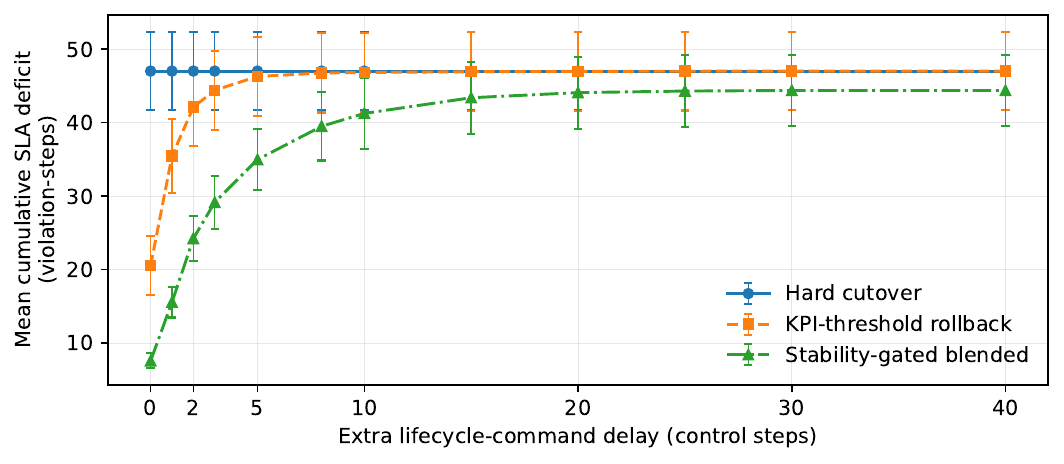}
\caption{Corrective lifecycle performance as command delay increases. Points
show means over 150 independently trained candidate policies after averaging
20 matched noise realizations within each candidate; error bars are 95\%
confidence intervals across candidate-level means. KPI-threshold rollback
rapidly approaches hard cutover, while stability-gated blending degrades
more gradually. Control steps are dimensionless and should not be read as
milliseconds.}
\label{fig:delay}
\end{figure*}

The policies, regime shift, environment, and activation point remain fixed
while only corrective commands are delayed. Delay postpones KPI-threshold
rollback and negative blending-weight back-off for stability-gated blending; hard
cutover is unaffected. The blending gate uses the one-step change of a
quadratic state-energy surrogate as a back-off heuristic rather than a
formal Lyapunov certificate. Each candidate is evaluated with 20 matched
noise realizations over 0--40 extra control steps. Replications are averaged
within candidate, leaving 150 candidate-level means as the independent unit
of analysis; Fig.~\ref{fig:delay} shows 95\% confidence intervals across
those means.

The primary metric is post-shift cumulative SLA deficit. With no added
delay, the mean deficit is 47.02 violation-steps for hard cutover, 20.53 for
KPI-threshold rollback, and 7.57 for stability-gated blending. At five
steps, threshold rollback has already risen to 46.29, while blending reaches
34.97. Blending degrades more gradually, reaching 41.26 at ten steps and
44.38 at 40 steps. Its mean cumulative advantage over hard cutover therefore
falls from 83.9\% at zero delay to 5.6\% at 40 steps.

The steady-state metric converges even faster. Hard cutover averages 1.004;
blending rises from 0.250 at zero delay to 0.946 at five steps, 0.990 at
eight, and 0.994 at ten. A paired bootstrap comparison no longer resolves a
steady-state benefit from eight steps onward, although that result should
not be read as proof of equivalence. The remaining cumulative advantage is
consistent with protection that occurs earlier in the transient.

The message is not a universal ``collapse threshold.'' It is that different
corrective strategies lose value at different rates as command delay grows.
KPI-threshold rollback loses nearly all numerical benefit within a few
intervals; blending is more tolerant but eventually retains only a small
cumulative advantage. Timing sensitivity is therefore a property of the
lifecycle mechanism itself, even before a physical protocol threshold is
assigned.

\subsection{Two-sided models: timing becomes consistency}
Release~20 introduces model pairing for the two-sided CSI-compression use
case~\cite{ref:rel20wid}. If the UE-side and network-side components change
state asynchronously, the delay window can leave the endpoints using a
mismatched pair. The decoder may then receive a representation from an
encoder other than the one with which it is intended to interoperate.

That turns timing into a consistency problem. A future fast path for
two-sided lifecycle control should therefore define what happens when the
pair cannot transition coherently: both endpoints activate the intended
pair, or the procedure enters a known recovery state. A common pair or
transaction identifier, a target activation time, acknowledgements, and
deterministic timeout handling are possible building blocks. The exact
mechanism is a standards-design choice; these pair-consistency semantics are
our proposal, not a completed Release~20 procedure.

\subsection{Why tuning cannot replace a timing requirement}
Shorter monitoring periods, tighter thresholds, or preferential handling may
improve a particular implementation, but they do not answer the specification
question. Faster monitoring increases signalling overhead; tighter thresholds
trade false negatives for false positives; and prioritization cannot by
itself create an end-to-end guarantee.

The structural issue is simpler: switching and fallback can become
latency-critical when monitoring drives them automatically, yet the WIDs do
not state an end-to-end actuation bound. Parameter tuning can optimize a
system against a requirement; it cannot substitute for a requirement that
has not been specified. This is where 3GPP mobility provides a useful
precedent.

\section{LTM: A Useful Architectural Precedent}
\label{sec:ltm}
3GPP mobility already provides an example of separating rich configuration
from time-critical execution.

The Release~18 work item on Further NR Mobility Enhancements describes the
starting point directly: serving-cell change relied on Layer~3 measurements
and RRC signalling, with lower-layer reset and reconfiguration contributing
to latency, signalling overhead, and interruption time~\cite{ref:ltmwid}.
The resulting split is visible in the protocol specifications. TS~38.331
clause~5.3.5.18.1 configures one or more LTM candidates in the
\texttt{LTM-Config} IE. TS~38.321 clause~5.18.35 then allows the network to
trigger the cell switch with an LTM Cell Switch Command MAC CE, whose Target
Configuration ID selects the preconfigured candidate in clause~6.1.3.75~\cite{ref:ts38331,ref:ts38321}.
In short, configuration is prepared in advance and the time-critical
selection is compact.

The analogy to model lifecycle management is architectural, not literal.
LTM changes the serving radio link and must deal with physical-layer access
procedures; model activation does not. What the two cases share is a decision
pattern: alternatives can be prepared ahead of time, while the final
selection becomes more valuable when it can be executed quickly.

That pattern suggests a clean question for AI/ML lifecycle control. Existing
AI/ML inference, monitoring, applicability, and data-collection configuration
can remain in RRC, and future candidate state can remain in Layer~3. If a
measured lifecycle timing budget shows that switch, fallback, or rollback
cannot be executed with enough margin through the Layer~3 path, the final
selection could use a compact lower-layer trigger.

This is intentionally a reuse argument rather than a call for a second
configuration framework. The candidate configuration remains where it is;
only the time-critical trigger would need new semantics, together with any
small amount of transaction state required for paired transitions. That
approach is consistent with RP-250766's lean-and-streamlined principle~\cite{ref:lean}:
add fast-path machinery only where timing justifies it.

\begin{table*}[!t]
\caption{Proposed lifecycle event classes and control-plane treatment}
\label{tab:classes}
\centering
\footnotesize
\setlength{\tabcolsep}{3.2pt}
\renewcommand{\arraystretch}{1.16}

\begin{tabularx}{\textwidth}{
@{}
>{\raggedright\arraybackslash}p{2.55cm}
>{\raggedright\arraybackslash}p{2.05cm}
>{\raggedright\arraybackslash}p{1.25cm}
>{\raggedright\arraybackslash}p{1.85cm}
>{\raggedright\arraybackslash}p{3.05cm}
>{\raggedright\arraybackslash}X
@{}}
\toprule
\textbf{Event class} &
\textbf{Trigger source} &
\textbf{Latency} &
\textbf{Consistency} &
\textbf{Proposed control path} &
\textbf{Consequence of delay} \\
\midrule

Monitoring configuration update &
Network policy &
Relaxed &
None &
RRC &
Stale monitoring configuration \\

Initial model configuration / candidate-set update &
Network policy; model delivery &
Relaxed &
None &
RRC &
Delayed candidate availability \\

Planned activation of a validated model &
Network policy &
Moderate &
Single-node &
RRC &
Delayed benefit only \\

Model switch among pre-configured candidates &
L1/L3 monitoring metric &
Tight &
Single-node &
RRC-configured candidates; L2 trigger if timing requires &
Continued operation of a degraded model \\

Rollback to a previously validated version &
Stability or drift detector &
Tight &
Single-node &
RRC-configured candidates; L2 trigger if timing requires &
Excursion can become an SLA violation \\

Fallback to the non-AI baseline &
Monitoring failure or timeout &
Tightest &
Local fallback state &
Local fallback; protected L2 indication if needed &
Defined fallback may not be reached in time \\

Paired activation or rollback (two-sided) &
Monitoring at either node &
Tight &
Pair-consistent transition &
L2 trigger plus pair-consistency coordination if timing requires &
Model-pair mismatch or inconsistent state \\

\bottomrule
\end{tabularx}
\end{table*}

\section{Which Lifecycle Events Need a Fast Path?}
\label{sec:req}
A fast path is useful only if the event is actually time-sensitive.
Lifecycle actions differ in urgency, in the consequence of delay, and in
whether both endpoints must change state together. Table~\ref{tab:classes}
therefore separates configuration events from candidate fast-path events.

The latency labels in Table~\ref{tab:classes} are qualitative design classes,
not standardized 3GPP timing values. Three design rules follow.

\textbf{1) Specify timing by event class.} The upper rows are naturally
configuration-oriented. The lower rows are potentially latency-critical
only when they are used as corrective control actions. If 6G defines such a
trigger, its useful timing property should be specified with the procedure
rather than left entirely to implementation.

\textbf{2) Treat two-sided transition as a pair-consistency problem.}
Faster delivery shortens a mismatch window, but it does not eliminate the
case of a lost or asymmetric command. The protocol therefore needs a defined
pair outcome and recovery rule. A lightweight realization could combine a
pair or transaction identifier with a target activation time or
acknowledgement and deterministic timeout recovery.

\textbf{3) Keep fallback locally executable.} A remote switch or paired
transition may reasonably depend on acknowledgement or coordination. Local
fallback should not: it may be needed precisely when peer coordination has
failed. Each node should therefore retain a defined non-AI fallback that can
be entered on timeout or detected failure, while a remotely received
fallback trigger remains subject to the security treatment discussed in
the security discussion below.

These rules build on the Release~20 lifecycle primitives rather than
relabeling functionality that is already standardized. They identify the
additional properties that a fast corrective path would need: a timing class,
pair-consistency and recovery semantics, and a locally executable fallback.

\section{Securing a Layer-2 Trigger}
\label{sec:sec}
RAN2 has already raised the broader security question that such a trigger
would inherit. During the 6G study phase, R2-2507945 asked SA3 for early
alignment on access-stratum protection for lower-layer control information,
including which information is critical, whether integrity and/or ciphering
is required, and what overhead is acceptable~\cite{ref:ls}.

The liaison does not identify AI/ML lifecycle commands specifically. Here,
we treat lifecycle triggers as one possible future class of lower-layer
control information and ask what follows from the event classes in
Table~\ref{tab:classes}.

\subsection{The security boundary}
The existing boundary is clear. TS~38.323 clause~4.3.1 lists ciphering and
integrity protection among the services PDCP provides to its upper layers,
including RRC. TS~33.501 clauses~6.5.1 and~6.5.2 place RRC integrity and
confidentiality protection at PDCP, and clause~6.5.1 states that layers below
PDCP are not integrity protected~\cite{ref:ts38323,ref:ts33501}. A lifecycle
trigger placed below PDCP would therefore need an explicit protection model
rather than simply inheriting RRC security.

\subsection{Forgery is not one risk}
The consequence of a forged command depends on what the command does. A
forged fallback to a defined non-AI mode is primarily a denial-of-benefit
attack: the AI/ML feature is lost even though no unvalidated model is
selected. A forged switch or rollback can select a poor candidate from the
operator-configured set. A forged paired transition can be more disruptive
because it may leave the UE-side and network-side components using an
inconsistent model pair.

The monitoring path deserves the same attention. An attacker who falsifies
the observation feeding a lifecycle decision may cause the network to issue
an authentic but incorrect switch, fallback, or retention decision. A future
security study should therefore consider both the trigger and the monitoring
input that causes it.

\subsection{Availability matters too}
For corrective lifecycle control, suppressing a valid command can be as
important as forging one. A blocked rollback or fallback may leave a
degraded model active beyond its useful intervention window, and integrity
protection alone does not prevent suppression. This is why remote-command
protection and local timeout behavior should be designed together.

A remotely received fallback trigger can still require integrity protection,
because forged fallback can deny the AI/ML benefit. At the same time, a node
should retain a locally executable fallback on timeout or loss of peer
coordination, so failure of a remote security exchange cannot prevent entry
to a defined non-AI state.

\subsection{A practical protection profile}
R2-2507945 deliberately leaves the protection choice open~\cite{ref:ls}. For
lifecycle triggers, integrity is the minimum property that should be
evaluated because an unauthenticated switch can directly alter network
behavior. Confidentiality should depend on what the trigger reveals: a
compact identifier selecting among already configured candidates may have
different needs from a report that carries sensitive measurements. Freshness
is equally important. A valid old trigger must not be replayable, so the
trigger should be bound to anti-replay or transaction context through
existing access-stratum security state or an explicit sequence mechanism.

Taken together, the control-loop view adds two requirements to that threat
model: monitoring reports that drive lifecycle decisions must be protected
appropriately, and local fallback must remain possible when protected remote
coordination cannot complete. These are topics for SA3/RAN2 study, not claims
about a protection profile already standardized for 6G.

\section{What 6G Needs to Decide}
\label{sec:open}
The 6G study phase is the right time to make the timing distinction explicit.
Once lifecycle procedures and protocol allocation are fixed, adding a timing
class or a lower-layer trigger becomes much more disruptive.

Four questions deserve early treatment.

\textbf{Which lifecycle events actually need a timing bound?} Configuration
updates can remain relaxed; monitoring-driven switch, rollback, fallback,
and paired transition may not. The qualitative classes in
Table~\ref{tab:classes} are a starting point, not standardized timing values.

\textbf{Can the existing Layer~3 path meet those bounds with adequate
margin?} A Layer~2 trigger should follow from a measured need, not from an
assumption that lower is always better. The first step is to establish a
physical actuation budget and compare it with feasible RRC and lower-layer
procedure delays.

\textbf{How should two-sided transitions preserve pair consistency?} A
specified target pair, completion rule, and timeout recovery path are needed
if independently supplied endpoints can change state at different times.

\textbf{What happens when coordination itself fails?} A node should retain a
locally executable non-AI fallback even when a protected remote transition
cannot complete. The corresponding remote trigger, if one exists, still
needs integrity, freshness, and an event-appropriate confidentiality policy.

The delay sweep makes the first two questions concrete. With no added delay,
stability-gated blending reduces mean cumulative SLA deficit from 47.02 to
7.57 violation-steps relative to hard cutover. At 40 extra control steps,
that deficit rises to 44.38. KPI-threshold rollback loses its benefit even
faster. Because the simulator does not map control steps to milliseconds and
does not measure RRC or MAC-CE latency, these results cannot set a 6G timing
requirement. They do establish the dependency that the standards discussion
must address: corrective lifecycle performance changes materially with
actuation delay.

A sensible standards sequence therefore starts with timing, not protocol.
Quantify the physical intervention window for the latency-critical lifecycle
classes; determine whether the existing Layer~3 path meets it; and introduce
a compact lower-layer trigger only if the margin is insufficient. If such a
trigger is needed, LTM offers a concrete configuration/trigger pattern, while
two-sided operation adds pair-consistency and recovery requirements and
failure handling requires a locally executable fallback.

\section{Biographies}
\vspace*{-30pt}
\begin{biographynophoto}
{DHARMENDRA KUMAR} [Member, IEEE] (dharmendra.sharma4@ieee.org) is an independent researcher based in Dallas, Texas, USA. His research interests include 5G/6G RAN, Cloud RAN, O-RAN, AI-native RAN control, reinforcement learning for network automation, and AI/ML lifecycle management.
\end{biographynophoto}

\end{document}